
\magnification=1200
\hoffset=-.1in
\voffset=-.2in
\vsize=7.5in
\hsize=5.6in
\tolerance 10000
\pretolerance = 5000
\baselineskip 12pt plus 1pt minus 1pt


\def \d{\displaystyle }
\def \t{\textstyle}
\def \jj{{\cal J}}

\def\ick{\eqalignno}

\def\overrightarrow#1{\vbox{\ialign{##\crcr
    \rightarrowfill\crcr\noalign{\kern-1pt\nointerlineskip}
    $\hfil\displaystyle{#1}\hfil$\crcr}}}
\def\overleftarrow#1{\vbox{\ialign{##\crcr
    \leftarrowfill\crcr\noalign{\kern-1pt\nointerlineskip}
    $\hfil\displaystyle{#1}\hfil$\crcr}}}

\def\ick{\eqalignno}
\def\sqr#1#2{{\vcenter{\hrule height.#2pt
   \hbox{\vrule width.#2pt height#1pt \kern#1pt
       \vrule width.#2pt}
         \hrule height.#2pt}}}

\def\_{^{}_}

\pageno=0
\centerline {\bf GAUGE FORMULATION OF THE SPINNING BLACK HOLE}
\centerline {{\bf IN (2+1)-DIMENSIONAL ANTI-DE SITTER SPACE}\footnote{*}{This
work is supported in part by funds
provided by the U. S. Department of Energy (D.O.E.) under contract
\#DE-AC02-76ER03069, by the Natural Sciences and Engineering
Research Council of Canada (M.L. and R.B.M.) and by the Swiss National
Science Foundation (D.C.).}}
\vskip 24pt
\centerline {\bf D. Cangemi$^1$, M. Leblanc$^1$, and R.B. Mann$^2$ }
\vskip 12pt

\centerline{\it $^1$Center for Theoretical Physics}
\centerline{Laboratory for Nuclear Sciences}
\centerline{and Department of Physics}
\centerline{Massachussetts Institute of Technology}
\centerline{Cambridge, Massachussetts 02139 U.S.A.}
\vskip 12pt

\centerline{$^2$Department of Physics}
\centerline{University of Waterloo}
\centerline{Waterloo, Ontario, Canada, N2L 3G1}
\vfill
\centerline{\bf ABSTRACT}
\medskip
We compute the group element of SO(2,2) associated with the spinning black
hole found by Ba\~nados, Teitelboim and Zanelli in (2+1)-dimensional anti-de
Sitter space-time. We show that their metric is built with
SO(2,2) gauge invariant quantities and satisfies Einstein's equations
with negative cosmological constant everywhere except at $r=0$.
Moreover, although the metric is singular on the horizons, the group
element is continuous and possesses a kink there.
\vfill
\centerline{Submitted to: {\it Physical Review Letters\/}}
\vskip 36pt
\noindent WATPHYS-TH92/09

\noindent CTP\#2162\hfill November 1992
\eject

\baselineskip 24pt plus 2pt minus 2pt

\noindent
The desire to have an interesting and mathematically tractable
setting for studying quantum gravity can in part be fulfilled by
investigating the properties of lower dimensional black holes [1].
Recently interest in this subject has increased because of the discovery by
Ba\~nados, Teitelboim and Zanelli~[2] (BTZ) of a spinning black hole
solution in $(2+1)$ dimensions. The aim of this note is to associate an
element of the anti-de Sitter group with this solution and to discuss its
physical  characteristics.

We use a Chern-Simons formulation of gravity~[3] with the gauge field
$$
{\cal A} =e^aP_a + \omega^aJ_a \eqno (1)
$$
decomposed in a basis of the Lie algebra so(2,2)
$$
[J_a, J_b]=\epsilon_{ab}{}^c J_c \quad ,\qquad
[J_a, P_b]=\epsilon_{ab}{}^c P_c \quad , \qquad
[P_a, P_b]={1\over l^2} \epsilon_{ab}{}^c J_c  \eqno (2)
$$
[$a,b,c=0,1,2$; $\epsilon^{012}=1$ and indices are lowered
with the metric $\eta_{ab}= {\rm diag} (-1,1,1)$].
In the absence of matter, the equations of motion
$$
{\cal F} =d{\cal A} + {\cal A} \wedge {\cal A} = 0\eqno (3)
$$
imply the Einstein's equations with negative cosmological constant
$\Lambda=-l^{-2}$ provided we identify
$e^a_\mu $ with the {\it Dreibein}
($g_{\mu\nu}=e^a_\mu e^b_\nu \eta_{ab}$)
and $\omega_\mu^a$
with the spin-connection
[$de^a+\epsilon^a_{\;\;bc}\,\omega^b{\wedge}~e^c=0$ is obtained from Eq.~(3)].
When Eq.(3) holds, the gauge field is pure gauge
$$
{\cal A}=U^{-1}dU \eqno(4)
$$
given by an element $U$ of the SO(2,2) group.

Here we shall obtain the group element $U$ associated with the
spinning black hole metric of Ref.~[2]
$$
ds^2_{\rm BTZ}=-N^2 dt^2 + N^{-2}dr^2+r^2(N^{\phi}dt+ d\phi)^2  \eqno (5)
$$
$$
\quad -\infty<t<\infty\,,\quad 0<r<\infty\,,\quad
0\leq \phi\leq 2\pi$$
with lapse and angular shift functions
$$\ick{
N^2(r)&=-M+{r^2\over l^2}+{J^2\over 4r^2}={1\over l^2}{1\over r^2}\left (
r^2-r_+^2\right )\left ( r^2 - (\jj r_+)^2\right )\quad  \cr
N^\phi(r)&=-{J\over 2r^2}=-\jj l\left ({r_+\over l}\right )^2
{1\over r^2}\quad .&(6) \cr }
$$
where
$\jj={Jl\over 2r_+^2}$ and $r_+^2={Ml^2\over 2}\left \{ 1 +
\left [ 1 -\left({J\over Ml}\right )^2\right ]^{1/2}\right \}$. Here, $r_+$ and
$r_-=|\jj| r_+$ are the two
zeroes for the lapse function $N(r)$. Of these, $r_+$ is the outer
horizon, which exists only if the requirements
$M>0$, $|J|\leq Ml\,$ (equivalent to $|\jj|<1$) hold ; this is henceforth
assumed. We
shall see that the metric (5) is locally anti-de Sitter
{\it i.e.}, it satisfies locally the Einstein's equation with negative
cosmological constant.
In fact, using the following change of complex variables
$$
{t'}= i r_+ \left( {t\over l} - \jj
\phi  \right )\quad , \qquad
{r'}=-il\,{\sqrt {\nu^2 (r)}} \quad , \qquad
{\phi'}=-i{r_+\over l} \left( \jj {t\over l} - \phi
\right ) \eqno (7)
$$
with $\nu^2(r)=1+{(r/r_+)^2 - 1\over 1-\jj^2}$ and taking real values for
$(t',r',\phi')$,
the above metric is globally transformed in the anti-de Sitter form
$$
ds^2_{\rm adS} = -\left[1+\left({r'\over l}\right)^2\right] dt'^2 +
\left[1+\left({r'\over l}\right)^2\right]^{-1} dr'^2 + r'^2 d\phi'^2 \eqno(8)
$$
with a wedge removed~[4], $(t',r',\phi')\equiv (t'-2\pi\jj r_+,r',\phi'+
2\pi r_+/l)$. [Note that the Jacobian of this transformation is singular
at $r=0$ and $r= r_-$.] We shall use this correspondence to find
the BTZ group element from the anti-de Sitter one.

In the anti-de Sitter case (8),
the {\it Dreibein} and the spin-connection may be chosen
(up to a Lorentz transformation) as
$$\vbox {\settabs 2 \columns
\+$e^0=\sqrt {1+(r'/l)^2}\, dt'$ & $\omega^0=-\sqrt {1+(r'/l)^2}\, d\phi '$\cr
\+$\d e^1={1\over \sqrt {1+(r'/l)^2}}\, dr'$& $\omega^1=0  $ \cr
\+$e^2=r'\,d\phi '$ & $\omega^2=-r'/l^2\, dt ' $ . \cr }\eqno (9)
$$
To find $U$, we have to solve Eq.~(4); explicitly
$$\ick {
\sqrt {1+(r'/l)^2 }\, P_0 - r'/l^2\, J_2 &= U^{-1}\partial_{t'}U \cr
{1\over\sqrt {1+(r'/l)^2 }}\, P_1 &= U^{-1}\partial_{r'}U &(10) \cr
r'\,P_2 -\sqrt {1+(r'/l)^2 }\, J_0 &= U^{-1}\partial_{\phi '}U\quad . \cr}
$$
Employing the {\it Ansatz} (inspired by the Poincar\'e
case)
$$
U= e^{\Theta^0J_0} e^{\Omega^0P_0} e^{\Omega^1P_1} e^{\Omega^2 P_2}\quad
\eqno (11)
$$
where $\Theta^0$, $\Omega^a$ are functions of $(t', r', \phi ')$, and,
making use of the commutation relations (2),
the right hand side of Eq.~(10) can be put in closed form yielding
eighteen differential equations.
These have the solution
$$
U_{\rm adS}(t',r',\phi ')=\exp \{-\phi ' {J_0}\}\,
\exp \{t' {P_0}\}\, \exp \{l\, {\sinh^{-1}} (r'/l)\, {P_1}\} \eqno (12)
$$
which is unique up to a left-multiplication by an invertible constant
matrix, reflecting the fact that we are dealing with first order
differential equations.

Since Eq.~(4) is covariant under coordinate transformations,
performing the analytic change of variables (7) on the
{\it Dreibein}, the spin-connection and the group element of the
anti-de Sitter case will give us a new version of these quantities
that still obey Eq.~(4) and a new metric,
which coincides with the spinning black hole metric (5). We have to
distinguish the three regions $r > r_+$, $r_- < r < r_+$
and $0< r <  r_-$.

Let us first consider $r > r_+$.  We need to specify how the function
$\sqrt {1+ (r'/l)^2}$ is defined when $r'/l\to -i\,\nu(r)$; this function
appears explicitly in (9) and implicitly in (12) [since
$\sinh^{-1}(r'/l) = \log ( (r'/l)+\sqrt {1+(r'/l)^2})$].
Our specification is
$$
\sqrt {1+(r'/l)^2}\;\;\to\;\;
-i\sqrt {\nu^2(r)-1} \quad\quad {\rm for} \quad r > r_+ \quad . \eqno (13)
$$
Then from Eq. (9), we get [$\nu^2(r)>1$ and $\nu(r)=\sqrt {\nu^2(r)}$
for $r>r_+$]
$$\vbox {\settabs 2 \columns
\+$\d e^0_{\rm BTZ}= r_+\sqrt {\nu^2(r)-1}
\left({dt\over l} - \jj d\phi\right)$ &
$\d\omega^0_{\rm BTZ}=-{r_+\over l}\sqrt{\nu^2(r)-1}
\left(d\phi - \jj{dt\over l} \right) $\cr
\+$\d e^1_{\rm BTZ}={l\over r_+^2}
{1\over (1-\jj^2)}{r\over\nu (r)\sqrt{\nu^2(r)-1}}dr$ &
$\omega^1_{\rm BTZ}= 0 $ \cr
\+$\d e^2_{\rm BTZ}=r_+\nu(r)\left ( d\phi - \jj {dt\over l} \right) $ &
$\d\omega^2_{\rm BTZ}=-{r_+\over l}\nu (r)\left({dt\over l} -
\jj d\phi \right)$ \cr }\eqno (14)
$$
and
$$\ick {
&U(t,r {>} r_+,\phi)=U_{\rm adS}((t'(t,r,\phi), r'(t,r,\phi),
\phi '(t,r,\phi)) & (15) \cr
&\t=\exp \left\{i{r_+\over l}\left(\jj {t\over l}-\phi \right ) \,{J_0}
\right \}\,\exp \left\{ir_+\left ({t\over l}-\jj \phi \right)\,{P_0}\right \}\,
\exp \{l\, \log\bigl(-i\nu(r) -i\sqrt{\nu^2(r)-1}\bigr )\,{P_1}\}
\quad .\cr}
$$
One can check that Eq.~(14) reproduces the metric (5) and that Eq.~(15)
is indeed a solution of Eq.~(4) with the components of ${\cal A}$ given
by (14). However, it does not belong to the group
but to its complexification.
This problem is easily dealt with by left-multiplying the element (15)
by $\exp\left ( {i{\pi\over 2}l \; P_1}\right )$, since
a solution of Eq.~(4) is determined up to a left
multiplication by a constant matrix,
$$\ick {
&\t U_{\rm BTZ}(t,r{>}r_+,\phi)= \t \exp \left
\{i{\pi\over 2}l \; P_1\right \} \; U(t,r{>}r_+,\phi) \cr
&=\t \exp \left \{-{r_+\over l}\left ({t\over l} -\jj \phi \right)\,{J_2}
\right \}\,
\exp \left \{-r_+\left(\jj {t\over l}-\phi \right ) \,{P_2}\right \}\,
\exp \left \{l\, {\cosh^{-1}} \nu(r)\,P_1\right \} &(16) \cr}
$$
which now belongs to the group. [We choose the positive branch of
$\cosh^{-1}$.]

Let us now turn to the second region, taking $ r_-<r<r_+$.
In this case the lapse function $N^2(r)$ changes sign,
characteristic of an horizon, which
interchanges the role of two coordinates. Thus, we get a real
{\it Dreibein} and a real spin-connection from the anti-de Sitter case by
interchanging $(e^0,\omega^0)$ and $(e^1,\omega^1)$
and we obtain [$0<\nu^2(r)<1$ and $\nu(r)=\sqrt{\nu^2(r)}$
for $ r_-<r<r_+$]
$$\vbox {\settabs 2 \columns
\+$\d e^0_{\rm BTZ}={l\over r_+^2}
{1\over (1-\jj^2)}{r\over \nu (r)\sqrt{1-\nu^2(r)}}dr$ &
$\omega^0_{\rm BTZ}= 0 $ \cr
\+$\d e^1_{\rm BTZ}= -r_+\sqrt {1-\nu^2(r)}
\left({dt\over l} - \jj d\phi\right)$ &
$\d\omega^1_{\rm BTZ}={r_+\over l}\sqrt{1-\nu^2(r)}
\left(d\phi - \jj{dt\over l} \right) $\cr
\+$\d e^2_{\rm BTZ}=r_+\nu(r)\left ( d\phi - \jj {dt\over l} \right) $ &
$\d\omega^2_{\rm BTZ}=-{r_+\over l}\nu (r)\left({dt\over l} -
\jj d\phi \right)$\quad . \cr }\eqno (17)
$$
As before we get from Eq.~(12) a solution to Eq.~(4)
$$\ick {
&U_{\rm BTZ}(t, r_-{<} r {<} r_+ ,\phi)\cr
&=\t
\exp \left\{-{r_+\over l}\left ({t\over l}-\jj \phi \right)\,J_2\right\}
\exp \left\{-r_+\left(\jj {t\over l}-\phi \right ) \,P_2\right\}\,
\exp \left\{-l\,\cos^{-1} \nu(r)\,P_0\right \} \quad .&(18) \cr}
$$
At the horizon, the group elements $U_{\rm BTZ}$ given by Eqs.~(16) and (18)
coincide, but the gauge field $\cal A$
has a discontinuity at $r=r_+$ as can be seen
from Eqs.~(14) and (17). Moreover, the {\it Dreibein} is degenerate
on the horizon since one of its components is zero and
another one diverges.

Finally, for $0< r < r_-$, the lapse function changes sign again, yielding
a metric signature similar to that outside of the outer horizon.
The transformation (7)
of the gauge field components (9), after a cyclic permutation, gives
[$\nu^2(r)<0$ and $|\nu(r)|=\sqrt{-\nu^2(r)}$ for $0<r< r_-$]
$$\vbox {\settabs 2 \columns
\+$\d e^0_{\rm BTZ}=-r_+ |\nu(r)|\left ( d\phi - \jj {dt\over l} \right) $ &
$\d\omega^0_{\rm BTZ}={r_+\over l}|\nu (r)|\left({dt\over l} -
\jj d\phi \right)$\cr
\+$\d e^1_{\rm BTZ}= -r_+\sqrt {1-\nu^2(r)}
\left({dt\over l} - \jj d\phi\right)$ &
$\d\omega^1_{\rm BTZ}={r_+\over l}\sqrt{1-\nu^2(r)}
\left(d\phi - \jj{dt\over l} \right) $\cr
\+$\d e^2_{\rm BTZ}={l\over r_+^2}
{1\over (1-\jj^2)}{r\over |\nu (r)|\sqrt{1-\nu^2(r)}}dr$ &
$\omega^2_{\rm BTZ}= 0 $\quad  \cr }\eqno (19)
$$
which satisfy $g_{\mu\nu}=e^a_\mu\,e^b_\nu\,\eta_{ab}$
and $de^a+\epsilon^a_{\;\;bc}\,\omega^b{\wedge}~e^c=0$ for this region.
We get the corresponding group element from Eq.~(12) and left multiplying
by $\exp \left \{ -{\pi\over 2} l \,P_0\right \}$
$$\ick {
U_{\rm BTZ}(t,0{<} r {<} r_- ,\phi)=&
\t
\exp \left\{-{r_+\over l}\left ({t\over l}-\jj \phi \right)\,J_2\right\}\,
\exp \left\{-r_+\left(\jj {t\over l}-\phi \right ) \,P_2\right\}\,\cr
&\t\times\,\exp \left\{-{\pi\over 2}l\, P_0\right \}\,
 \exp \left\{-l\,\sinh^{-1} |\nu(r)|\,P_2\right \} \quad .&(20) \cr}
$$
As before this solution and the previous one, Eq.~(18), match at $r=r_-$
although ${\cal A}$ does not.

With Eqs.~(16), (18) and (20), the general solution is
$$
U(x) = C\, U_{\rm BTZ}(x)\, \exp \{\lambda^a(x)J_a\} \eqno (21)
$$
where $C$ is a
matrix of integration constants, which is in the
group since $U(x)$ must be in the group, and the right
multiplication by  $\exp \{\lambda^a (x) J_a\} $
arises because the {\it Dreibein} and the spin-connection are
defined up to a local Lorentz transformation.
This solution is continuous in all space-time.
The existence of $U(x)$ insures that ${\cal F}=0$
({\it i.e.,} the metric (5) fulfills the anti-de Sitter Einstein
equations) almost everywhere, except maybe at
$r=0$, $r=r_-$ and $r=r_+$ where $U(x)$ is not differentiable.
A convenient way to see a nonvanishing
$\cal F$ is to use the Wilson loops. At the same time, this enables
us to compute the gauge invariant quantities.

Let us look at the Wilson loop enclosing the origin at fixed time
$$
W={\cal P} \exp \oint \, A_\mu \, dx^\mu \eqno (22)
$$
which is related to $U$ by
$$
W = U^{-1}(t,r,\phi=0) \, U(t,r,\phi=2\pi) \equiv e^w \quad .\eqno (23)
$$
A gauge transformation is equivalent to a conjugation of $W$
by a group element $V$
$$
W\rightarrow V\,W\, V^{-1} = e^{V w V^{-1}} \quad . \eqno (24)
$$
It is known how to extract quantities invariant under (24).
Consider the Casimir of the adjoint representation
of SO(2,2)
$$
{\cal C}= c_1 \left (J_aJ^a \,+\, l^2\,P_aP^a \right )\,
+\,c_2 \left (J_aP^a \,+\, P_aJ^a \right )\, \eqno (25)
$$
where $c_1\, , c_2$ are arbitrary constants. If $w=\theta^a\,J_a
\,+\, \xi^a\,P_a $ in Eq.~(24), it follows from the Casimir that
the two quantities
$$\ick {
m&= {1\over 2\pi^2}\left (\theta_a\theta^a +{1\over l^2}\xi_a\xi^a
\right ) \cr
j&= {1\over 2\pi^2} \theta_a\xi^a &(26) \cr }
$$
are invariant under (24).

The solution (21) gives to the Wilson loops (22) located in one of
the three regions considered a unique value up to a conjugation by a
group element depending on the position of the loop
$$
W=V(r)^{-1}\,\exp\left \{2\pi {r_+\over l}\left (\jj \, J_2 + l\, P_2\right )
\right\}\, V(r)\quad . \eqno(27)
$$
Eq.~(26) reveals that the first invariant quantity
coincides with $M$ and the second one with $J$, the constants parametrizing
the black hole metric (5). Thus $M$ and $J$ are physically relevant,
and according to~[5] correspond to the mass and spin of the black hole.

A small Wilson loop around the origin enables us to compute $\cal F$ at this
point. Namely for small $\varepsilon$
$$\ick{
W_{r=\varepsilon}&=1 +
\int_0^{\varepsilon} \,dr\,\int_0^{2\pi}\,d\phi\,
{\cal F}_{r\phi} + {\cal O}(\varepsilon^2) \cr
\log W_{r=\varepsilon}
&=2\pi {r_+\over l}V(\varepsilon)^{-1} \left (\jj \, J_2 + l\, P_2 \right)
V(\varepsilon) & (28) \cr}
$$
where the second equality is obtained from Eq.~(27). Since ${\cal F}=0$
outside $r=0$, it must be the distribution (we use the explicit expression
of $V(0)$, which is read off Eq.~(20))
$$
{\cal F}_{r\phi}=\delta(r)\,{r_+\over l}{1\over\sqrt{1-\jj^2}}
\left[J_1 - \jj^2\,J_0 + \jj l\,(P_1 - P_0)\right]
\quad . \eqno (29)
$$
In (2+1)-dimensions, the $J_a$ components of $\cal F$ are used to
build the Einstein tensor with cosmological constant.
For the {\it Dreibein} of Eq.~(19), Einstein's equations are
modified at the origin by the presence of the localized spinning black
hole
$$\ick {
\sqrt {-g} \left (G_{tt}+{1\over l^2} g_{tt}\right )
&= \left ({r_+\over l}\right )^4
\left (1+\jj^2\right )^2 \, \delta (r) &(30a) \cr
\sqrt {-g}\left (G_{t\phi}+{1\over l^2}  g_{t\phi}\right )
&=-\left ({r_+\over l}\right )^4
(1+\jj^2)\,{\jj l} \, \delta (r) &(30b) \cr
\sqrt {-g}\left (G_{\phi\phi}+{1\over l^2}  g_{\phi\phi}\right )
&= \left ({r_+\over l}\right )^4
\, {(\jj l)^2} \,  \delta (r) &(30c) \cr }
$$
without any modification to the other components.

Similarily, the $P_a$
components of $\cal F$ provide the torsion tensor and we find here that
the torsion free condition is modified at the origin by
$$\ick {
de^0 + \epsilon^0{}_{ab}\,\omega^a \wedge e^b &=
-{\jj \,r_+\over {\sqrt {1-\jj^2}}}
\,\delta (r) \,dr\wedge d\phi &(31a) \cr
de^1 + \epsilon^1{}_{ab}\,\omega^a \wedge e^b &=
{\jj \,r_+\over {\sqrt {1-\jj^2}}}
\,\delta (r) \,dr\wedge d\phi &(31b) \cr }
$$
and no modification for the other component. The fact that the Wilson
loops (27) belong always to the same class [under the conjugation
(24)] indicates the absence of matter at the horizons $r=r_\pm$,
extending the validity of Eqs.~(30,31) to the whole space-time.
The RHS's correspond to the
energy-momentum tensor of the matter, which is responsible for this
particular geometry. This is in contrast with the (3+1)-dimensional
Schwarschild metric configuration, which cannot be written as the
solution of Einstein's equations with a source localized at the origin.
The difference is understood from the absence of gravitational interaction in
(2+1) dimensions. We stress that Eqs.~(30,31) are valid only for the
choice of {\it Dreibein} (19). For example, a rotation of the {\it Dreibein} by
$\exp \phi\,J_0$ would add to the RHS of Eq.~(29) the term $\delta (r) \, J_0$.

It is interesting to comment on the Poincar\'e case $l\to \infty$
({\it i.e.,} a vanishing cosmological constant). In this limit and
for $M>0, \, J\neq 0$, the metric (5)
becomes
$$
ds^2 = (-\sqrt{M}dt+{J\over{2\sqrt{M}}}d\phi)^2  +
{{dr^2}\over{{J^2\over{4r^2}}-
M}} + (r^2 - {J^2\over{4M}}) d\phi^2 \eqno(32)
$$
and is a solution
of the corresponding limit equations (30), {\it i.e.,} Einstein's
equations with a singularity at the origin.  Near the source
(for $r<{J\over{2\sqrt{M}}}$), the space-time has closed
timelike curves, $\phi$ being a periodic time coordinate.
For  $r>{J\over{2\sqrt{M}}}$ the space-time is
a helix expanding in the time coordinate $r$,
since the variable $R\equiv -\sqrt{M}t + {J\over{2\sqrt{M}}}\phi$
increases by an amount
$\pi{J\over{\sqrt{M}}}$ whenever $\phi\rightarrow\phi+2\pi$.
For $J=0$  the metric  becomes
$$
ds^2 = -{1\over M} dr^2 + M dt^2 +r^2 d\phi^2 \eqno (33)
$$
which corresponds to a cylinder expanding in the time coordinate $r$.

The gauge formulation of gravity is useful in the analysis of
classical solution. We have computed the group element associated
with the BTZ black hole. We can choose it continuous but not
differentiable where the metric does not exist. We have shown that the
gauge invariants coincide with the geometric ADM invariants [5].
Using the Wilson loops, we were able to locate the source
of matter at the origin appearing in the anti-de Sitter Einstein's equations.

\vskip 12pt
\noindent{\bf Acknowledgements}
This study was suggested by Roman Jackiw. We thank him as well as
Roger Brooks, Michael Crescimanno and Miguel Ortiz for helpful discussions.
\vskip 6pt

\vskip 12pt
\noindent{\bf References}
\vskip 6pt

\item{1.} For a review see R.B. Mann, {\it Gen. Rel. Grav.} {\bf 24},
433 (1992).
\medskip
\item{2.} M.~Ba\~nados, C.~Teitelboim and J.~Zanelli, {\it Phys.
Rev. Lett.} {\bf 69}, 1849 (1992).
\medskip
\item{3.} A.~Ach\'ucarro and P.~Townsend, {\it Phys. Lett.} {\bf 180B},
89 (1986); E.~Witten, {\it Nucl. Phys.} {\bf B311}, 46 (1988/89).
\medskip
\item{4.} S.~Deser, R.~Jackiw and G.~'t~Hooft, {\it Ann. Phys.} (NY)
{\bf 152}, 220 (1984); S.~Deser and R.~Jackiw, {\it Ann. Phys.} (NY) {\bf
153}, 405 (1984).
\medskip
\item{5.} M.~Ba\~nados, M. Henneaux and C. Teitelboim, in preparation.

\end